\documentclass[aps,twocolumn,prd]{revtex4}
\input epsf
\usepackage{graphicx}% Include figure files
\newcommand{\beq}{\begin{equation}}
\newcommand{\beqa}{\begin{eqnarray}}
		  \newcommand{\eeq}{\end{equation}}
\newcommand{\eeqa}{\end{eqnarray}}

\newcommand{\lsim}{\lesssim}
\newcommand{\gsim}{\gtrsim}

\newcommand{\vect}[1]{\mbox{\boldmath${#1}$}}
\newcommand{\lmk}{\left(}
\newcommand{\rmk}{\right)}
\newcommand{\lnk}{\left\{ }
\newcommand{\rnk}{\right\} }
\newcommand{\lkk}{\left[}
\newcommand{\rkk}{\right]}
\newcommand{\lla}{\left\langle}

\newcommand{\rra}{\right\rangle}

\newcommand{\vex}{{\vect x}}

\newcommand{\vey}{\vect y}

\begin{document}
%\baselineskip 8mm

%%%%%%%%%%%%%%%%%%%%%%%%%%%%%%%%%%%%%%%%%%%%%%%%%%%%%%%%%%%%%%%%%%%%%%%%%%
%%%%%%%%%%%%%%%%%%%%%%%%%%%%%%%%%%%%%%%%%%%%%%%%%%%%%%%%%%%%%%%%%%%%%%%%%%
\title{ Demagnified GWs from Cosmological Double Neutron Stars and GW Foreground Cleaning Around 1Hz} 
%%%%%%%%%%%%%%%%%%%%%%%%%%%%%%%%%%%%%%%%%%%%%%%%%%%%%%%%%%%%%%%%%%%%%%%%%%
%%%%%%%%%%%%%%%%%%%%%%%%%%%%%%%%%%%%%%%%%%%%%%%%%%%%%%%%%%%%%%%%%%%%%%%%%%
%
%
%
%%%%%%%%%%%%%%%%%%%%%%%%%%%%%%%%%%%%%%%%%%%%%%%%%%%%%%%%%%%%%%%%%%%%%%%%%%
\author{Naoki Seto}
%%%%%%%%%%%%%%%%%%%%%%%%%%%%%%%%%%%%%%%%%%%%%%%%%%%%%%%%%%%%%%%%%%%%%%%%%%
\affiliation{Department of Physics, Kyoto University
Kyoto 606-8502, Japan
}
%%%%%%%%%%%%%%%%%%%%%%%%%%%%%%%%%%%%%%%%%%%%%%%%%%%%%%%%%%%%%%%%%%%%%%%%%%
\date{\today}
%
%
%
%
%
%
%%%%%%%%%%%%%%%%%%%%%%%%%%%%%%%%%%%%%%%%%%%%%%%%%%%%%%%%%%%%%%%%%%%%%%%%%%
\begin{abstract}
 Gravitational waves (GWs) from cosmological double neutron star binaries (NS+NS) can be significantly demagnified by strong gravitational lensing effect, and the proposed future missions such as BBO or DECIGO might miss some of the  demagnified GW signals  below a detection threshold. 
The undetectable binaries would form a GW foreground which might hamper detection of a very weak primordial GW signal. We discuss the outlook of this potential problem, using a simple model based on the singular-isothermal-sphere lens profile.  Fortunately, it is expected that, for presumable merger rate of NS+NSs, the residual foreground would be below the detection limit $\Omega_{GW,lim}\sim 10^{-16}$ realized with BBO/DECIGO by correlation analysis.

\end{abstract}
\maketitle

%%%%%%%%%%%%%%%%%%%%%%%%%%%%%%%%%%%%%%%%%%%%%%%%%%%%%%%%%%%%
%%%%%%%%%%%%%%%%%%%%%%%%%%%%%%%%%%%%%%%%%%%%%%%%%%%%%%%%%%%%
\section{introduction}
%%%%%%%%%%%%%%%%%%%%%%%%%%%%%%%%%%%%%%%%%%%%%%%%%%%%%%%%%%%%
%%%%%%%%%%%%%%%%%%%%%%%%%%%%%%%%%%%%%%%%%%%%%%%%%%%%%%%%%%%%
A GW background from the early universe is one of the primary targets of observational cosmology.
 It would provide us with crucial information on  physics at  very high-energy scales.  Among others, a primordial GW background generated during an inflationary period is a key objective \cite{Allen:1996vm}. Currently, there are two approaches to probe the inflation background. One is  indirect observation around $f\sim 10^{-18}$Hz through  B-mode polarization of cosmic microwave background \cite{Seljak:1996gy}. Another is  direct GW detection around $f\sim 0.1$-1Hz with the proposed space laser interferometers such as the Big Bang Observer (BBO) \cite{bbo} or the Deci-hertz Interferometer Gravitational Wave Observatory (DECIGO) \cite{Seto:2001qf,Kawamura}. These two approaches at widely separated frequencies are complimentary and we might disclose  fundamental properties of inflation by using them simultaneously (see {\it e.g.} \cite{Smith:2005mm}).  But, in both of them, we must cope with strong  astrophysical contaminations to uncover the inflation background. 

At present,  the overall profile of astrophysical  GW foregrounds  around $f\sim1$Hz is unclear. Although  white-dwarf binaries  (important for the Laser Interferometer Space Antenna (LISA) \cite{lisa}) would not make a critical limit  there \cite{Farmer:2003pa}, we might have a strong foreground component whose quantitative properties are difficult to  predict now. This is partly due to the complicated astrophysical processes involved.  For example, to estimate the foreground made by supernovae of  population III stars, we need, at least,   the formation rate of these stars and their angular momentum distribution.  But, unfortunately,  these important elements are poorly known at present \cite{Buonanno:2004tp}. Meanwhile, around 1Hz, we also have a foreground made by double neutron star binaries (NS+NSs) that are individually very simple system accurately predicted with the post-Newtonian inspiral waveforms \footnote{This is also true for compact binaries with stellar-mass black holes (BHs). But we do not discuss them in this paper, since NS+BHs or BH+BHs have larger chirp masses than NS+NSs and would be detected more easily.}. In addition, NS+NSs are promising target for the first detection of GWs with ground-based detectors, and their coalescence rate has been extensively discussed with the observed abundance of NS+NSs in our galaxy \cite{Kalogera:2001dz}. Considering our current understanding on the 1Hz band, in this paper, we  study only the NS+NS foreground, neglecting other potential but highly uncertain astrophysical foregrounds.

{ NS+NSs can be regarded as solid GW sources around 1Hz, and  cleaning of these binaries would be a critical element for the success of BBO/DECIGO to detect a weak primordial GW background. }
The basic approach for the cleaning is to identify individual binaries and subtract their chirping waveforms from the data of detectors \cite{Cutler:2005qq,Harms:2008xv}.
As we see later, in order to fully use the designed sensitivity of detectors for a primordial background,  the residual astrophysical foreground after the cleaning must be $\sim 10^5$ times smaller  than their original strength (in terms of  GW energy spectrum).

To  discuss the prospect of this cleaning procedure, we first need to understand the detectability of GWs from individual NS+NSs.  Since we cannot expect a large fluctuation ({\it e.g.} a factor of 2) for the intrinsic chirp mass distribution of NS+NSs, the primary parameter that characterizes the signal-to-noise ratio of a binary would be its distance or equivalently its redshift $z$.  Another important parameter of a binary is its inclination angle,  which can change the GW amplitude  by a factor of $2\sqrt{2}$ (ratio between  face-on and  edge-on binaries).   Therefore,  the basic requirement for the cleaning is to make detectors that have enough sensitivities to detect an edge-on NS+NS at high redshift ({\it e.g.} $z=10$) \cite{Cutler:2005qq}.

However, the situation becomes complicated due to the gravitational lensing effect that modulates  observed amplitudes of GWs during their propagation between the sources and detectors \cite{lens}.  We might miss  some demagnified  GW signals which are below a detection threshold, and their resultant residuals might be an obstacle for detecting a weak inflation background.  Our principle aim in this paper is to provide a rough outlook on this potential problem caused by gravitational lensing. 

Gravitational lensing effects can be  broadly divided into two categories: the weak lensing due to accumulated small distortions during  wave propagation \cite{Bartelmann:1999yn} (see also \cite{Cutler:2009qv,Itoh:2009iy} for GWs from NS+NSs) and the strong lensing caused by specific massive objects with large distortions \cite{lens}.  We are interested in significantly  ({\it e.g.} 50\%) demagnified lensing fluctuations, but such  probability is known to be negligible for weak lensing, even for a high-redshift source \cite{Barber:2000pe}. Therefore, we  concentrate on the strong lensing effect that can generate significant demagnification more frequently.  For the lens profile, we use the singular isothermal sphere (SIS) model, which is fairly successful for studying various observational aspects of strong lensing effects \cite{lens}.

This paper is organized as follows. In Sec. II, we discuss the merger rate of cosmological NS+NSs and their GW foreground. In Sec. III, the strong gravitational lensing effect is studied. The probability distribution function  for the faint-end of demagnification is evaluated with the SIS model.  Then we estimate the GW foreground made by undetectable NS+NSs. Section IV is devoted to  discussions on this paper.

%%%%%%%%%%%%%%%%%%%%%%%%%%%%%%%%%%%%%%%%%%%%%%%%%%%%%%%%%%%%
%%%%%%%%%%%%%%%%%%%%%%%%%%%%%%%%%%%%%%%%%%%%%%%%%%%%%%%%%%%%
\section{cosmological NS+NS binaries}
%%%%%%%%%%%%%%%%%%%%%%%%%%%%%%%%%%%%%%%%%%%%%%%%%%%%%%%%%%%%
%%%%%%%%%%%%%%%%%%%%%%%%%%%%%%%%%%%%%%%%%%%%%%%%%%%%%%%%%%%%

In this section, we briefly discuss basic aspects of cosmological NS+NSs and their GWs around 1Hz, following Cutler and Harms \cite{Cutler:2005qq}.
First we evaluate the total merger rate ${\dot N}_{T}$  ($T$, total) of cosmological NS+NSs. Based on the observed NS+NSs in our galaxy and the number densities of various types of nearby galaxies, the comoving merger rate  at present $ {\dot n}_0$ is estimated to be $10^{-8}$-$10^{-6}{\rm Mpc^{-3}yr^{-1}}$, roughly corresponding to the Advanced-LIGO detection rate $10^{1\pm1}{\rm yr^{-1}}$ \cite{Kalogera:2001dz}.  
To deal with merger events at cosmological distances, we put the  comoving merger rate $ {\dot n}(z)$ at redshift $z$ as follows
\beq
{\dot n}(z)=s(z)  {\dot n}_0.
\eeq
  Here, the nondimensional function $s(z)$ represents the redshift dependence of the rate, and, in this paper, we consider the following two concrete models: (I) the fiducial evolutionary model $s_I(z)$ with a piecewise function  
\begin{eqnarray}
s_I(z)=\left\{ \begin{array}{ll}
1+2z &   (0\le z \le  1)  \\
3(5-z)/4   &   (1< z \le 5)  \\
0 & (5<z), \\
\end{array} \right.
\end{eqnarray} 
 and (II) a simple model  
\begin{eqnarray}
s_{II}(z)=\left\{ \begin{array}{ll}
1 &   (0\le z \le  10)  \\
0 & (10<z) \\
\end{array} \right.
\end{eqnarray} 
with a flat merger  rate   up to $z=10$. The first model $s_I$ is the same as that used in Cutler and Harms \cite{Cutler:2005qq}.

 For estimating the total merger rate ${\dot N}_{T}$ observed today,  we need  to calculate the comoving volume element.
The comoving  distance to an object at redshift $z$ is given by 
\beq
r(z)=c\int_0^z \frac{dz}{H(z)},\label{comd}
\eeq
where
$H(z)$ is the Hubble parameter at redshift $z$. In this paper, we fix the cosmological parameters at $H_0=70{\rm km~ sec^{-1} Mpc^{-1}}$, $\Omega_m=0.3$ and $\Omega_\Lambda=0.7$ for a flat universe.   The function $H(z)$ is written by
\beq
H(z)=H_0 \sqrt{\Omega_m(1+z)^3+\Omega_{\lambda}}.
\eeq

  Since the comoving volume of a shell  between   $z$ and $z+dz$ is given by  $4\pi r(z)^2 (dr/dz) dz$,  the total merger rate  is expressed as \cite{Cutler:2005qq}
\beq
{\dot N}_{T}=4\pi c \int_0^\infty dz \frac{r(z)^2  {\dot n}(z)}{(1+z) H(z)} \label{rate}
\eeq
with the redshift factor $(1+z)^{-1}$ due to the cosmological time dilution.  Here,  we neglected a tiny increase of the actual number of  merger signals due to the multiple lensed signals.  For the fiducial model $s(z)=s_I(z)$, we numerically obtain
\beq
{\dot N}_{T}=1.1\times 10^5 \lmk \frac{{\dot n}_0}{\rm 10^{-7}Mpc^{-3}yr^{-1}}\rmk {\rm yr^{-1}}. \label{ar}
\eeq
  For the flat-rate model $s(z)=s_{II}(z)$, the prefactor in Eq.(\ref{ar}) becomes $5.9\times 10^4$.    In Fig. 1, we show the relative redshift distribution of the merger event ${\dot N}(>z)/{\dot N}_{T}$.  The numerator of this ratio is defined by
\beq
{\dot N}(>z)=4\pi c \int_z^\infty dz \frac{r(z)^2  {\dot n}(z)}{(1+z) H(z)}.
\eeq

Next we evaluate the spectrum of the  total GW energy $\Omega_{GW,BT}$ ($B$, binary; $T$, total) emitted by the chirping  NS+NSs before their subtractions.   Here, following the standard convention, the energy density $\Omega_{GW,BT}$  is defined  per the logarithmic frequency interval and  normalized by the critical density of the universe.
Its formal expression is given by \cite{Phinney:2001di}
\beq
\Omega_{GW,BT}(f)=\frac{8\pi^{5/3} G^{5/3}M_c^{5/3} f^{2/3}}{9c^2 H_0^2}\int_0^\infty dz \frac{{\dot n}(z)}{(1+z)^{4/3} H(z)}. \label{omgbt}
\eeq
  { In Eq.(\ref{omgbt}) the frequency $f$ is measured at the detector frame, and  $M_c$ is  the intrinsic (not redshifted) chirp mass at the source frame. We keep using these two definitions throughout this paper.}  We also fix the chirp mass at $M_c=1.22 M_\odot$.
 The redshift dependence of the integral 
${\dot n}(z)/[(1+z)^{4/3}H(z)]$ in Eq.(\ref{omgbt}) can be decomposed into two factors: the event rate proportional to 
\beq
r(z)^2 {\dot n}(z)/[(1+z)H(z)]
\eeq
 as given in Eq.(\ref{rate}), and the square of the individual signals 
\beq
h^2\propto 1/[r(z)^2(1+z)^{1/3}]
\eeq
 (see discussions later in this section).  For our fiducial model $s(z)=s_I(z)$, we numerically obtain
\beq
\Omega_{GW,BT}(f)=4.0\times 10^{-12}  \lmk \frac{{\dot n}_0}{\rm 10^{-7}Mpc^{-3}yr^{-1}}\rmk  \lmk \frac{f}{\rm 1Hz}\rmk^{2/3},
\eeq
while the prefactor becomes $2.1\times 10^{-12}$
 for $s(z)=s_{II}(z)$.

%%%%%%%%%%%%%%%%%%%%%%%%%%%%%%%%%%%%%%%%%%%%%%%%%%%%%%%%%%%%%%%%%%%%%%%%%%
%%%%%%%%%%%%%%%%%%%%%%%%%%%%%%%%% Figure 1 %%%%%%%%%%%%%%%%%%%%%%%%%%%%%%%
%%%%%%%%%%%%%%%%%%%%%%%%%%%%%%%%%%%%%%%%%%%%%%%%%%%%%%%%%%%%%%%%%%%%%%%%%%
\begin{figure}
  \begin{center}
\epsfxsize=7.cm
\begin{minipage}{\epsfxsize} \epsffile{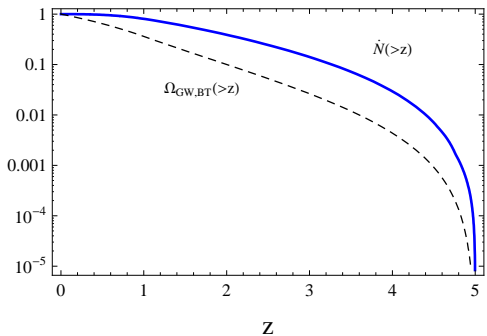} \end{minipage}
 \end{center}
  \caption{Fraction of the observed merger rate ${\dot N}_T$ and the foreground amplitude $\Omega_{GW,BT}$ by NS+NSs more distant than a given redshift. We plot numerical results for the fiducial  merger rate $s(z)=s_I(z)$ (see also \cite{Cutler:2005qq}).
 }
%\label{f2}
\end{figure}
%%%%%%%%%%%%%%%%%%%%%%%%%%%%%%%%%%%%%%%%%%%%%%%%%%%%%%%%%%%%%%%%%%%%%%%%%%
%%%%%%%%%%%%%%%%%%%%%%%%%%%%%%%%%%%%%%%%%%%%%%%%%%%%%%%%%%%%%%%%%%%%%%%%%%

The correlation analysis is a powerful method to detect weak stochastic GW signals \cite{Flanagan:1993ix}.  In terms of the normalized energy density $\Omega_{GW}$, we can, in principle, realize a factor of $\sim SN^{-1}(f T_{obs})^{1/2}$  ($SN$, the signal-to-noise ratio for detection;  $T_{obs}$, observation period) better sensitivity, compared with the sensitivity $\Omega_{GW,detector}\sim 10^{-14}$ directly corresponding to the noise spectrum of BBO. With BBO/DECIGO and an observational period of $T_{obs}\sim 10$yr, the limiting sensitivity of the correlation analysis to a stochastic  GW background is $\Omega_{GW,lim}\sim 10^{-16}$ around the optimal frequency  $f\sim 0.3$Hz where we are free from the potential foreground by double white-dwarf binaries \cite{Seto:2005qy}. { This limiting sensitivity $\Omega_{GW,lim}$ is a not a simple power-law function of  GW frequency $f$, because of the shapes of   the overlap reduction function (for definition see {\it e.g.} \cite{Flanagan:1993ix}) and the detector noise spectrum.} 
In order to fully exploit the potential specification of the proposed detectors  and to pursue a weak primordial GW background  down to their limiting sensitivity $\Omega_{GW,lim}\sim 10^{-16}$, we need to subtract the NS+NS foreground  and make the residual 
smaller by at least a factor of  
\beq
\frac{\Omega_{GW,lim}}{\Omega_{GW,BT}}\sim 5.6\times 10^{-5}\lmk\frac{{\dot n}_0}{10^{-7}{\rm Mpc^{-3} yr^{-1}}}\rmk^{-1} \label{tar}
\eeq
around the optimal frequency of BBO/DECIGO $f\sim 0.3$Hz,  by identifying  NS+NSs up to their highest redshift.  Note again that the above target residual level ($\Omega_{GW,lim}\sim 10^{-16}$) is much lower than the detector noise level (corresponding to $\Omega_{GW,detector}\sim 10^{-14}$).
In Fig. 1, we show the fraction of the foreground $\Omega_{GW,BT}$  made by binaries more distant than a given redshift $z$.

Now we discuss the observation of chirping GWs from individual NS+NSs  with BBO/DECIGO.
Here, we study typical binaries using the unperturbed GW amplitudes.  The lensed signals would be analyzed in Sec.III.
 In what follows, we neglect dependence on sky positions  and polarization angles of binaries, and we apply the averaged response of detectors with respect to these parameters.    { This is because  multiple detectors with different orientations would be used for BBO/DECIGO \cite{bbo,Kawamura}, and we can expect a relatively weak dependence of  signal-to-noise ratio on the direction and polarization angles, owing to an effective averaging effect (see {\it e.g.} \cite{Seto:2004ji} for averaging on the direction angles). This prescription simplifies our  analysis (see also \cite{Cutler:2005qq}). But,  with only one detector, more detailed studies would be required for sources with short signal durations ({\it e.g.} 1 week).}

We can evaluate   the amplitudes of the two polarization modes of a  binary with the quadrupole formula. In the frequency domain and in the principle polarization coordinate, their explicit forms for a circular orbit are given by \cite{Thorne_K:1987}
\beq
(h_+,h_\times)=\sqrt{\frac5{96}}\frac{\pi^{-2/3} G^{5/6} M_c^{5/6} f^{-7/6}}{(1+z)^{1/6}c^{3/2}r(z)} (1+u^2,2u).
\eeq
   Here,  we defined the geometrical parameter $u\equiv \cos I$  using the inclination angle $I$.   A face-on  (edge-on) binary has  $I=0$ ($I=\pm \pi/2$ respectively). Meanwhile, for a NS+NS at frequency $f$,  the time $T_{GW}$ before the merger  is given by
\beq
T_{GW} \sim1 \lmk \frac{f}{0.2{\rm Hz}}\rmk^{-8/3} (1+z)^{-5/3}~{\rm yr},
\eeq
 and  smaller than  planned operation period of BBO/DECIGO.

For signal analysis of each binary, we assume that, with some workable methods,  the residual foreground would  eventually become  smaller than the detector noise after  subtractions of binary signals \cite{Cutler:2005qq}.  This justifies our treatment below  in which we  consider only the detector noises for estimating the signal-to-noise ratio of individual binary.  Note that this is a much weaker assumption compared with the previous requirement for realizing the ultimate sensitivity for  detecting the weak GW background with correlation analysis.  We  put the signal-to-noise ratio $\rho (z,u)$ of   unperturbed chirping GW from a  NS+NS at a redshift $z$ as follows
\beq
\rho(z,u)=\frac{B M_c^{5/6} F(u)^{1/2}}{(1+z)^{1/6} r(z)}
\eeq
with a function
$F(u)\equiv1+6u^2+u^4$ (see \cite{Cutler:2005qq}).  
Here,
$B$ is a constant determined by  the noise spectrum of the detectors. Given the designed sensitivity of BBO, we fix the parameter $B$ so that the signal-to-noise ratio becomes  $\rho(z=5,u=0)=40$ to match   \cite{Cutler:2005qq}.  { For other redshifts,  we have $\rho(z=10,0)=30, \rho(3,0)=52$ and $\rho(1,0)=110$. 
    We also have $\rho(z,1)/\rho(z,0)=(F(1)/F(0))^{1/2}=2\sqrt{2}$  for dependence on the inclination angle, as commented in Sec. 1.} 
For the fiducial merger model $s=s_I(z)$ with a realistic normalization ${\dot n}_0\lsim 10^{-6}{\rm Mpc^{-3} yr^{-1}}$, Cutler and Harms \cite{Cutler:2005qq} showed that the designed sensitivity of BBO [corresponding to $\rho(z=5,0)=40$] would enable us to detect all of  the (unlensed) binaries and make their residual below the limiting sensitivity $\Omega_{GW,lim}\sim 10^{-16}$.

For an evaluation later in Sec. IV, we also define the mean signal-to-noise ratios  by 
\beq
{\bar \rho}\equiv \frac{4\pi c \int_0^\infty dz \int_0^1 du \frac{r(z)^2  {\dot n}(z)}{(1+z) H(z)} \rho(z,u)^2}{{\dot N}_T}.
\eeq
 We numerically evaluated this expression and obtained ${\bar \rho}=187$ for $s=s_{I}$ and ${\bar \rho}=161$ for $s=s_{II}$.

%%%%%%%%%%%%%%%%%%%%%%%%%%%%%%%%%%%%%%%%%%%%%%%%%%%%%%%%%%%%
%%%%%%%%%%%%%%%%%%%%%%%%%%%%%%%%%%%%%%%%%%%%%%%%%%%%%%%%%%%%
\section{demagnified GW signals}
%%%%%%%%%%%%%%%%%%%%%%%%%%%%%%%%%%%%%%%%%%%%%%%%%%%%%%%%%%%%
%%%%%%%%%%%%%%%%%%%%%%%%%%%%%%%%%%%%%%%%%%%%%%%%%%%%%%%%%%%%
 
%%%%%%%%%%%%%%%%%%%%%%%%%%%%%%%%%%%%%%%%%%%%%%%%%%%%%%%%%%%%
\subsection{Lensing Probability}
%%%%%%%%%%%%%%%%%%%%%%%%%%%%%%%%%%%%%%%%%%%%%%%%%%%%%%%%%%%%

Strong gravitational lensing produces multiple signals for an intrinsically single event \cite{lens}.  In this subsection, we discuss the faint end of GW signals caused by strong lensing. For the density profile of the lens objects,  we use the SIS  that captures  lens structures well at the scales relevant for the strong lensing events \cite{lens}.  Here, we do not include detailed effects such as the external tidal shear field or the ellipticity of  lenses, which cause minor corrections for the strong lensing probability itself (see {\it e.g.} \cite{Huterer:2004jh}, and discussions on the quadruple lenses therein).

The explicit form of the SIS density profile $\frac{v^2}{3\pi G R^2}$  ($R$: the radial coordinate) is characterized by the one-dimensional velocity dispersion $v$.   We further  introduce the unperturbed  angular diameter distances $D_{os}$, $D_{ol}$ and $D_{ls}$ between the observer-source, observer-lens and lens-source respectively. For example, the distance $D_{ol}$ is given by 
\beq
D_{ol}=\frac{r(z_l)}{1+z_l}
\eeq
  with the comoving distance $r(z)$ defined in Eq.(\ref{comd}).  In order to study lens mapping, we  define the characteristic angle 
\beq
\theta_d\equiv \frac{4\pi v^2 D_{ls}}{c^2 D_{os}} \label{td}
\eeq
 and use the two-dimensional coordinate $\vey$  for the source plane normalized with the length unit $\theta_d D_{os}$. We also introduce  the  coordinate  $\vex$ for  the lens plane normalized with the length  $\theta_d D_{ol}$.
The origin of these two coordinates are fixed at the direction of the lens center from the observer. 
 Then the lens equation for the SIS model is given by  \cite{lens}
\beq
\vey=\vex-\vex/|\vex| . \label{le}
\eeq
Because of  the apparent symmetry around the lens center, the lens mapping between $\vex$ and $\vey$ is  essentially one-dimensional correspondence on  lines passing their origins.
  The lens equation (\ref{le}) has two solutions 
\beq
x_\pm=y\pm 1
\eeq
 for a small impact parameter at $0\le y<1$, and one solution 
\beq
x=y+1
\eeq
 for a large impact parameter at $y\ge 1$.  

The amplification factor $A$ of gravitational lensing is evaluated with the Jacobian of the mapping  and given by
\beq
A=\sqrt{\left| \frac{x}y \frac{dx}{dy}\right|}. \label{ampa}
\eeq
For the two solutions $x_\pm$,  the amplification (\ref{ampa})  becomes 
\beq
A_+=\sqrt{1/y+1},~~~A_-=\sqrt{1/y-1}.
\eeq
Comparing these two signals, 
 the second one  corresponds to the inner, fainter signal with a later arrival time.   Note that the amplification $A$ is defined by the square root of the magnification $\mu$ which  is often used in the literature on gravitational lensing of electro-magnetic waves  \cite{lens}. Our definition here reflects the important  fact that, for detecting GW from a binary, we directly observe the waveform itself ($\propto h$), not its energy ($\propto h^2$).

In contrast to the identity $A_+>1$, 
the fainter counterpart $A_-$ approaches  zero in the limit $y\to 1$ (from below). Since we are   interested in weak GW signals that might be undetectable, we mainly study   the fainter one $A_-$ rather than the brighter  one $A_+$.  The signal-to-noise ratio of a demagnified signal is expressed as  $A_{-}\rho(z,u)$.

 At $y\sim 1$,  the time delay between two signals is given by \cite{lens}
\beqa
\Delta t&\sim& 32\pi^2 \lmk \frac{v}c \rmk^4 \frac{D_{ol}D_{ls}}{cD_{os}} (1+z_l)\nonumber\\ &\sim& 8.6\times10^6 \lmk  \frac{v}{210 \rm km/sec}\rmk^4 \lmk  \frac{D_{ol}D_{ls} D_{os}^{-1}(1+z_l)}{\rm 1Gpc} \rmk  {\rm sec}.\nonumber
\eeqa
For  a typical velocity dispersion $v$ of a lens object,   we have the relation  $f\Delta t\gsim 10^6\gg 1$  around the optimal frequency of BBO/DECIGO $f\sim0.3$Hz.  Therefore, in the present situation, the geometrical optics approximation would work well \cite{Takahashi:2003ix}, and the  strongly-lensed two  signals would be observed as two distinct chirping waveforms. But it might be interesting to analyze potential wave effects ({\it e.g.} in relation to substructure of lenses) \cite{Takahashi:2003ix}. 
 For the SIS model, the separation angle $\Delta \theta$ between the two images  is given by \cite{lens}
\beq
\Delta \theta=2\theta_d\sim  2.4" \lmk \frac{D_{ls}}{D_{os}} \rmk  \lmk \frac{v}{\rm 210km/sec} \rmk^2. \label{dthe}
\eeq

With the expression for the characteristic angle $\theta_d$ defined in Eq.(\ref{td}), the probability of the strong lensing event for a source at redshift $z_s$
 is given by
\beq
P_d(z_s)=\int_0^{z_s} dz_l \int_0^{\infty} dv \frac{dN}{dv} \frac{16\pi^3 c^{-3} v^4 (1+z_l)^2 D_{ls}^2 D_{ol}^2}{H(z_l)D_{os}^2 }, \label{sp}
\eeq
where $\frac{dN}{dv}$ is the distribution function for velocity dispersions of lens objects. 
In this paper, we use the following model
\beq
v \frac{dN}{dv}=\frac{\phi_*  \beta}{\Gamma(\alpha/\beta)} \lmk \frac{v}{v_*}\rmk^\alpha \exp\lkk -\lmk \frac{v}{v_*}\rmk^\beta\rkk \label{vel}
\eeq
 characterized by parameters  $\phi_*=4.1\times 10^{-3}h^3{\rm Mpc^{-3}}$,  $v_*=88.8{\rm km ~sec^{-1}}$, $\alpha=6.5$ and $\beta=1.93$ \cite{Oguri:2007sv}.  We neglect  the cosmological evolution of this function.  This is   partially supported by actual  lensing observation up to $z_s\sim1$.  Our main results [{\it e.g.} Eqs.(\ref{r1}) and (\ref{r2})] change only slightly,  for an evolutionary model with the following set of parameters analyzed in \cite{Oguri:2007sv}
\beq
\phi_*(z_l)=\phi_*(1+z_l)^{-0.229},~~~\sigma_*(z_l)=\sigma_*(1+z_l)^{-0.01}.
\eeq
   The product $v^4 dN/dv$ is relevant for  the lensing probability and also for the distribution of  time delays. It becomes maximum at $v=2.4\sigma_*=210$km/sec and steeply declines at large $v$, as expected from the fraction 
\beq
\frac{\int_{4\sigma_*}^\infty dv  v^4 dN/dv}{\int_{0}^\infty dv  v^4 dN/dv}\sim 10^{-4}.
\eeq
  Even for a high-redshift source $z_s\sim 5$, the redshift integral (\ref{sp}) has dominant contribution around $z_l\sim 1$. In Fig. 2, we present the lensing probability $P_d(z)$, as a function of the source redshift $z$.

On the source plane, the demagnification $A_{-}(<1)$ corresponds to the normalized radial coordinate as
\beq
y=\frac1{A_{-}^2+1}.
\eeq
 Therefore, the probability that a source at a redshift $z$ has  a demagnification in the range $[A_{-},A_{-}+dA_{-}]$, is given by
\beq
\frac{dP}{dA_{-}} dA_{-}=P_d(z) \frac{4A_{-}}{(A_{-}^2+1)^3} dA_{-} \label{sisp}.\eeq

%%%%%%%%%%%%%%%%%%%%%%%%%%%%%%%%%%%%%%%%%%%%%%%%%%%%%%%%%%%%%%%%%%%%%%%%%%
%%%%%%%%%%%%%%%%%%%%%%%%%%%%%%%%% Figure 2 %%%%%%%%%%%%%%%%%%%%%%%%%%%%%%%
%%%%%%%%%%%%%%%%%%%%%%%%%%%%%%%%%%%%%%%%%%%%%%%%%%%%%%%%%%%%%%%%%%%%%%%%%%
\begin{figure}
  \begin{center}
\epsfxsize=8.9cm
\begin{minipage}{\epsfxsize} \epsffile{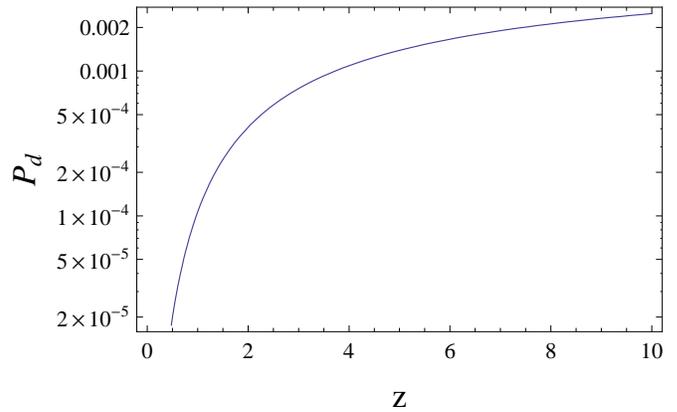} \end{minipage}
 \end{center}
  \caption{ Probability of having multiple lensed  images  as a function of the source redshift $z$.    For the velocity distribution function $dN/dv$, we use the no-evolution model given by Eq.(\ref{vel}).
 }
%\label{f2}
\end{figure}
%%%%%%%%%%%%%%%%%%%%%%%%%%%%%%%%%%%%%%%%%%%%%%%%%%%%%%%%%%%%%%%%%%%%%%%%%%
%%%%%%%%%%%%%%%%%%%%%%%%%%%%%%%%%%%%%%%%%%%%%%%%%%%%%%%%%%%%%%%%%%%%%%%%%%

%%%%%%%%%%%%%%%%%%%%%%%%%%%%%%%%%%%%%%%%%%%%%%%%%%%%%%%%%%%%
%%%%%%%%%%%%%%%%%%%%%%%%%%%%%%%%%%%%%%%%%%%%%%%%%%%%%%%%%%%%
\subsection{Undetectable Lensed Signals}
%%%%%%%%%%%%%%%%%%%%%%%%%%%%%%%%%%%%%%%%%%%%%%%%%%%%%%%%%%%%
%%%%%%%%%%%%%%%%%%%%%%%%%%%%%%%%%%%%%%%%%%%%%%%%%%%%%%%%%%%%

In this subsection, we study the  subtraction problem for lensed GWs emitted by cosmological NS+NSs. First we evaluate the event rate for strongly-lensed GWs. For binaries at redshift $z$, the fraction $R_0(z)$ of the strongly-lensed GWs coincides with the  probability $P_d(z)$ given in Eq.(\ref{sp})  as
\beq
R_0(z)\equiv P_d(z).
\eeq
Then the total merger rate of the lensed signals is written as
\beq
{\dot N}_{L}\equiv 4\pi \int_0^\infty dz \frac{r(z)^2 R_0(z)  {\dot n}(z)}{(1+z) H(z)}. \label{tl}
\eeq
For  the fiducial model $s=s_I(z)$, we numerically evaluated this expression and obtained the averaged lensing probability  
\beq
{\bar P}\equiv{\dot N}_{L}/{\dot N}_{T}=3.7\times 10^{-4}\label{19}
\eeq
  or equivalently 
\beq
{\dot N}_{L}=40 \lmk \frac{{\dot n}_0}{\rm 10^{-7}  Mpc^{-3}yr^{-1}}  \rmk {\rm yr^{-1}}.
\eeq
For the flat-rate model  $s(z)=s_{II}(z)$, we have the ratio  ${\dot N}_{L}/{\dot N}_{T}=8.6\times 10^{-4}$.

Next we discuss a demagnified GW signal whose  signal-to-noise ratio $A_{-}\rho(z,u)$ is less than a detection threshold $\rho_{th}$.  To begin with, we estimate the relevant value $\rho_{th}$, taking into account  the observational situation of the two lensed signals.   The typical time delay $\Delta t$ between the earlier bright signal with  $A_+$ and the later faint one with $A_-$ is much less than the planned observational period $T_{obs}\sim 10$yr of BBO/DECIGO.  Thus we mainly consider a scenario to search for the demagnified signals under  the detection of the bright ones \footnote{But note that this is not true in the initial phase of the observation.}.  Except for the coalescence time $t_c$, the fitting parameters of the normalized waveforms of the two lensed signals can be regarded as almost  identical, and these parameters would be generally well determined  with the bright first image. {   For example, with BBO,  the typical size of the localization error-ellipsoids in the sky is $\sim 10"\times 10"$ for a NS+NS at $z\sim 5$ \cite{Cutler:2005qq}.  On the other hand, the characteristic image separation $\Delta \theta$ in Eq.(\ref{dthe}) is several arcseconds.  Therefore,  directions of two lensed images would be fitted by  the same parameter for a NS+NS at $z\sim5$. But,  for a NS+NS at low redshift,  the second image could  get out of the error-ellipsoid of the brighter image.

 The similarities of fitting parameters would significantly ease the detection of the demagnified one with the matched filtering method, compared with its single detection for which we need $10^{30}$-$10^{36}$ templates to make a full coherent  signal integration.  Even with the estimated computational power available at the era of BBO/DECIGO, we cannot make so many templates, and we need to use a suboptimal detection method which requires  a higher detection threshold ($\rho_{cr}>20$ see \cite{Cutler:2005qq}) than  the (optimal) coherent integration.  Therefore, the information of the brighter signal considerably  decreases the detection threshold $\rho_{th}$ for a lensed second signal, compared with its single detection.  For the typical number of templates $n_{t_c}\sim10^7$-$ 10^8$ estimated from the binning of the  coalescence time $t_c$ of the faint image, the required condition for the threshold $\rho_{th}$ with respect to  a false alarm rate $f_r$ is given as \cite{Cutler:2005qq}
\beq
f_r\sim N_{t_c} {\rm erf}(\rho_{th}/\sqrt{2}),
\eeq
 and we have the solution $\rho_{th}\sim 6.5$ for $f_r\sim 0.01$.  
 Including a safety factor ({\it e.g.} potential resampling of other fitting parameters around the bright signal), we take $\rho_{th}=10$ as a standard value for detecting the faint second image.  For a reference we also use a pessimistic value $\rho_{th}\sim 20$ in the analyses below \cite{Cutler:2005qq}. The results for this higher value would also provide  us with  an insight about the stand-alone analysis for a demagnified signal without using the information of the associated  brighter one.

Here, we briefly comment on the identification of lensed pairs.  The order-of-magnitude estimation for the required numbers of binning for the redshifted chirp mass $M_c(1+z)$ is $\sim 10^8$, which is  much larger than the relevant numbers of NS+NSs $\sim 10^5$ \cite{Cutler:2005qq}. In addition, given the good localization in the sky,  directions of binaries would  also become  useful information to identify a lensed pair.  Therefore it is unlikely to misidentify GWs from two different NS+NSs as a lensed pair.  Although our aim in this paper is not to discuss further scientific possibilities with identified lensed pairs, such studies would be also interesting \cite{Cutler:2009qv,Itoh:2009iy}. }

Now  we  statistically study   the demagnified GW signals that are below the detection threshold with $A_{-}\rho(z,u)<\rho_{th}$.  The fraction   of the number of undetectable signals at a redshift $z$ is given as follows; 
\beq
R_1(z)\equiv \frac{\int_0^\infty dA_{-} \int_0^1 du \frac{dP}{dA_{-}} \Theta (\rho_{th}-A_{-}\rho(z,u))}{\int_0^1 du }.\label{r1}
\eeq
Here $\Theta(x)$ is the step function.
In Fig. 3, we plot the function $R_1(z)$ for  two choices: $\rho_{th}=10$ and 20. At the small amplification regime $A_{-}\ll 1$, the probability of having an  amplification less than a given value $A_{-}$ is proportional to $A_{-}^2$ [see Eq.(\ref{sisp})], and  we  have an  asymptotic scaling relation $R_1\propto \rho_{th}^2$.  But we should be careful to note that this scaling relation depends on the details of the lensing profile and is not universal.
The event rate ${\dot N}_{U}$ ($U$, undetectable) of the undetectable signals is evaluated as in Eq.(\ref{tl}) and we have
\beq
{\dot N}_{U}=4\pi \int_0^\infty dz \frac{r(z)^2 R_1(z)  {\dot n}(z)}{(1+z) H(z)}.
\eeq
For the fiducial model $s(z)=s_{I}(z)$ and the threshold $\rho_{th}=10$, we numerically obtained the result 
\beq
{\dot N}_{U}=1.1\lmk \frac{{\dot n}_{0}}{10^{-7}{\rm Mpc^{-3} yr^{-1}}}\rmk {\rm yr^{-1}},
\eeq   
 or equivalently 
\beq
{\dot N}_{U}/{\dot N}_{T}=1.1\times 10^{-5}.
\eeq
With the higher threshold $\rho_{th}=20$, the ratio becomes  ${\dot N}_{U}/{\dot N}_{T}=3.9\times 10^{-5}$.

We move to evaluate the energy  spectrum of the GW foreground made by undetectable binary signals with $A_{-}\rho(z,u)<\rho_{th}$.  Here we introduce the factor $R_2(z)$ as the fraction of the GW energy due to the undetectable signals at redshift $z$ relative to  the total binaries at the redshift.  Summing up the contribution of undetectable ones, the factor $R_2(z)$ is formally given by 
\beq
R_2(z)\equiv \frac{\int dA_{-} \int_0^1 du \frac{dP}{dA_{-}} \Theta (\rho_{th}-A_{-}\rho(z,u))\lnk A_{-}\rho(z,u)\rnk^2}{\int_0^1 du \rho(z,u)^2}.\label{r2}
\eeq
In Fig. 3 we provide the function $R_2(z)$ for two thresholds $\rho_{th}=10$ and 20.  In the present case, we can derive the asymptotic behavior $R_2\propto \rho_{th}^4$ for the SIS profile, after a simple consideration on the factors in the numerator of Eq.(\ref{r2}).

%%%%%%%%%%%%%%%%%%%%%%%%%%%%%%%%%%%%%%%%%%%%%%%%%%%%%%%%%%%%
%%%%%%%%%%%%%%%%%%%%%%%% Figure 3 %%%%%%%%%%%%%%%%%%%%%%%%%%
%%%%%%%%%%%%%%%%%%%%%%%%%%%%%%%%%%%%%%%%%%%%%%%%%%%%%%%%%%%%
\begin{figure}[ht]
\begin{center}
\epsfxsize=8.9cm
\epsffile{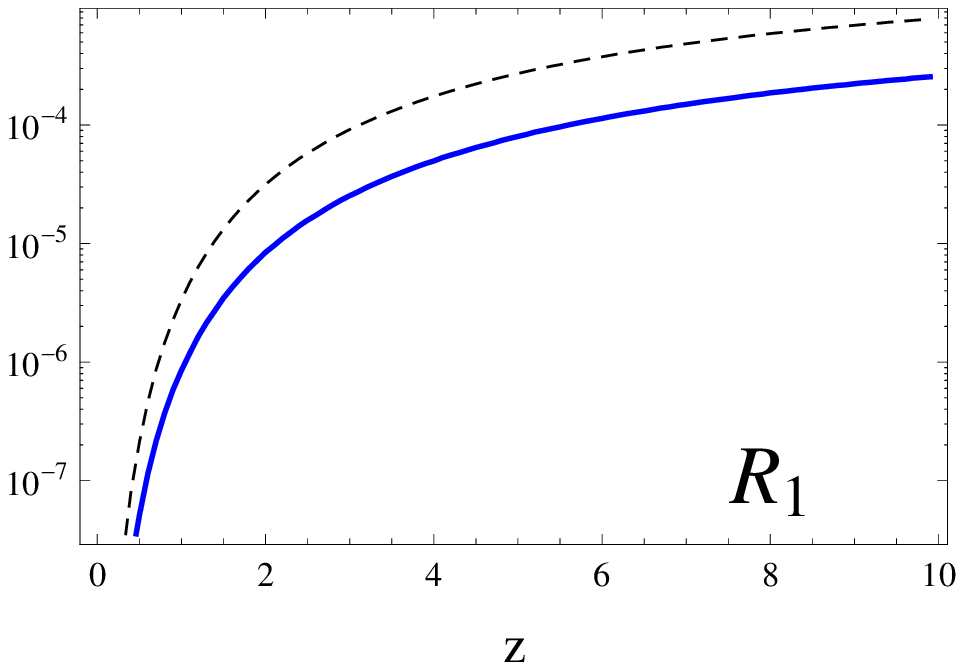}
\epsfxsize=8.9cm
\epsffile{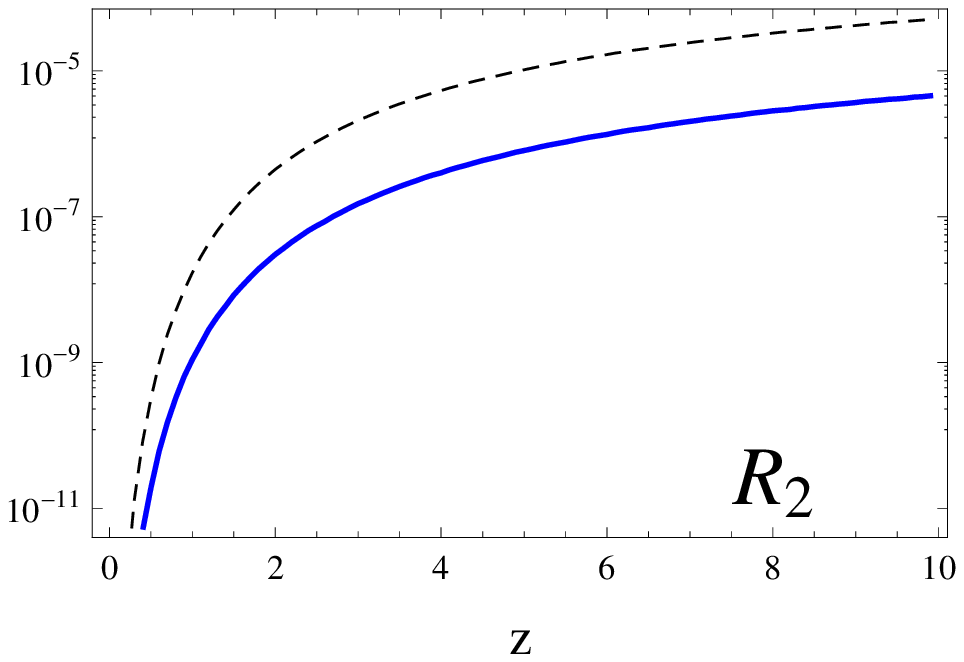}
\end{center}
\caption{  Relative contributions of the undetectable GW signals from NS+NSs at redshift $z$. The BBO noise curve  is assumed  with two choices of the detection threshold $\rho_{th}$ (dashed curve: $\rho_{th}=20$ and solid curve: $\rho_{th}=10$).
 Left panel: the number fraction $R_1(z)$ of the undetectable signals. Right panel: the fraction $R_2(z)$ of the GW energy density due to  the undetected  NS+NSs.} 
\end{figure}
%%%%%%%%%%%%%%%%%%%%%%%%%%%%%%%%%%%%%%%%%%%%%%%%%%%%%%%%%%%%
%%%%%%%%%%%%%%%%%%%%%%%%%%%%%%%%%%%%%%%%%%%%%%%%%%%%%%%%%%%%

With the fraction $R_2(z)$,  the energy spectrum $\Omega_{GW,BU}$ of the undetectable demagnified ones is expressed as \footnote{Note that this relation is valid for a fixed detector sensitivity. If we improve the sensitivity by a factor of $X$, we have  asymptotic  scaling relations $\Omega_{GW,BU} \propto \rho_{th}^4 X^{-4}$ and $\Omega_{GW,lim} \propto  X^{-2}$.}
\beq
\Omega_{GW,BU}\equiv \frac{8\pi^{5/3} M_c^{5/3} f^{2/3}}{9 H_0^2}\int_0^\infty \frac{{\dot n}(z)R_2(z)}{(1+z)^{4/3} H(z) }.
\eeq
With the fiducial model  $s(z)=s_I(z)$, we numerically obtained $\Omega_{GW,BU}/\Omega_{GW,BT}=1.5 \times 10^{-8}$ for $\rho_{th}=10$ and $\Omega_{GW,BU}/\Omega_{GW,BT}=2.1 \times 10^{-7}$ for $\rho_{th}=20$.  With $s(z)=s_{II}(z)$, we have $9.8\times 10^{-8}$ ($\rho_{th}=10$) and $1.2\times 10^{-6}$ ($\rho_{th}=20$).
Compared with the ratio in $(\ref{tar})$, the undetectable GW foreground would be comfortably smaller than the limiting sensitivity $\Omega_{GW,lim}\sim 10^{-16}$ for the realistic normalization ${\dot n}_0\lsim 10^{-6} {\rm Mpc^{-3} yr^{-1}}$.

%%%%%%%%%%%%%%%%%%%%%%%%%%%%%%%%%%%%%%%%%%%%%%%%%%%%%%%%%%%%
%%%%%%%%%%%%%%%%%%%%%%%%%%%%%%%%%%%%%%%%%%%%%%%%%%%%%%%%%%%%
\section{discussions}
%%%%%%%%%%%%%%%%%%%%%%%%%%%%%%%%%%%%%%%%%%%%%%%%%%%%%%%%%%%%
%%%%%%%%%%%%%%%%%%%%%%%%%%%%%%%%%%%%%%%%%%%%%%%%%%%%%%%%%%%%

So far, we have studied the demagnified GW signals  with  the SIS density profile.  Although this lens  model is quite simple, it has  reproduced   observational results (with electro-magnetic waves) of gravitational  lensing   fairly well \cite{lens}.  But one might think that  this  model has been discussed at the relatively high-amplification regime where  observations can be  performed more easily.
In this respect, it might be reasonable to wonder whether the simple SIS model is also a useful tool to study  the strong lensing effects at the low-amplification regime as discussed in this paper.  One simple example highlighting the situation is about the central third images of strong lensing \cite{Rusin:2000mm,Keeton:2002ed}. A lensing galaxy could have a core structure around its center, instead of the power-law  profile $\propto r^{-2}$ of the SIS model, and such a density profile can generate a very faint third image around the direction toward the center of the lens. However,   observational analysis of the third images is known to be quite difficult, partly due to the faintness of the signals \footnote{Fortunately the residual foreground by the third central images would not be a problem, given their typical magnitudes $\lla A_{-}^2\rra\lsim 0.001$ \cite{Keeton:2002ed}.}. In addition to this example, it should  be  mentioned that, compared with typical electro-magnetic wave observation, GWs are generally measured  at much lower frequencies and could be more susceptible to the wave effects ({\it e.g.} by substructures) that could complicate the signal analysis \cite{Takahashi:2003ix}.  In relation to the foreground cleaning which is essential for directly detecting a GW background from inflation, further studies on the faint end of lensing amplification would be worthwhile beyond our simple treatment.

On a final note, we attempt a more robust approach to evaluate the strength of the foreground composed by residual GW signals from undetected  binaries, without using details of the probability distribution (\ref{sisp}) for  the demagnification $A_-$.  Here, we simply assume that, if GW is strongly lensed, then the total signal-to-noise ratio of its unsubtracted component would be less than the detection threshold $\rho_{th}$.  We define the residual $\Omega_{GW,BS}$ ($S$; simplified) as the summation of these unsubtracted ones.
Since we have the relations $\Omega_{GW,BT}\propto N_T {\bar \rho}^2$ and  $\Omega_{GW,BS}\propto  N_L {\rho_{th}}^2=N_T {\bar P}{\rho_{th}}^2$ (see Eq.(\ref{19})), we obtain
\beq
\frac{\Omega_{GW,BS}}{\Omega_{GW,BT}}={\bar P}\lmk\frac{\rho_{th}}{\bar \rho}  \rmk^2.
\eeq
We regard the amplitude $\Omega_{GW,BS}$ as a conservative upper bound for the potential foreground made by the undetected signals.  With the  numerical results presented so far, we can  evaluate this ratio and obtain
\beq
\frac{\Omega_{GW,BS}}{\Omega_{GW,BT}}=1.1 \times 10^{-6}\lmk \frac{\rho_{th}}{10}\rmk^2
\eeq
 for the fiducial model $s(z)=s_{I}(z)$ and  $3.3 \times 10^{-6}(\rho_{th}/10)^2$ for $s(z)=s_{II}(z)$.  
Therefore, unless the normalization  ${\dot n}_0$ is relatively high ${\dot n}_0\gsim 10^{-6} {\rm Mpc^{-3} yr^{-1}}$, the upper limit  $\Omega_{GW,BS}$ is  smaller than the limiting sensitivity $\Omega_{GW,lim}$ [see Eq.(\ref{tar})], and we expect that the residual  foreground caused by demagnified signals would not be a fundamental problem to detect an inflation background of $\Omega_{GW}\gsim 10^{-16}$ with BBO and DECIGO.

%%%%%%%%%%%%%%%%%%%%%%%%%%%%%%%%%%%%%%%%%%%%%%%%%%%%%%%%%%%%%%%%%%%%%%%%%%%%%%%

The author would like to thank R. Takahashi for helpful discussions. 
This work was 
supported by Grants-in-Aid for Scientific Research of the Japanese Ministry of Education, Culture, Sports, Science,
and Technology Grant No. 20740151.

%%%%%%%%%%%%%%%%%%%%%%%%%%%%%%%%%%%%%%%%%%%%%%%%%%%%%%%%%%%%%%%%%%%%%%%%%%%%%%%
%%%%%%%%%%%%%%%%%%%%%%%%%%%%%%%%%%%%%%%%%%%%%%%%%%%%%%%%%%%%%%%%%%%%%%%%%%%%%%%
%%%%%%%%%%%%%%%%%%%%%%%%%%%%%%%%%%%%%%%%%%%%%%%%%%%%%%%%%%%%%%%%%%%%%%%%%%%%%%%
%\include{ref}
%%%%%%%%%%%%%%%%%%%%%%%%%%%%%%%%%%%%%%%%%%%%%%%%%%%%%%%%%%%%%%%%%%%%%%%%%%%%%%%

\end{document}